\documentclass[twocolumn,aps,prb,showpacs]{revtex4}

\usepackage{epsf}
\usepackage{bm}

\begin{document}

\title{%
Finite Temperature Behavior of Small Silicon and Tin Clusters:
An {\it Ab Initio} Molecular Dynamics Study.} 

\author{Sailaja Krishnamurty, Kavita Joshi, and D. G. Kanhere}

\affiliation{%
Department of Physics and
Centre for Modeling and Simulation,
University of Pune,
Ganeshkhind,
Pune--411 007,
India.}

\author{S. A. Blundell}

\affiliation{%
D\'epartement de Recherche Fondamentale sur la
Mati\`ere Condens\'ee, CEA-Grenoble/DSM \\
17 rue des Martyrs, F-38054 Grenoble Cedex 9, France.}

\date{\today}

\begin{abstract}

The finite temperature behavior of small Silicon (Si$_{10}$, Si$_{15}$,
and Si$_{20}$) and Tin (Sn$_{10}$ and Sn$_{20}$) clusters is studied
using isokinetic Born-Oppenheimer molecular dynamics. The lowest
equilibrium structures of all the clusters are built upon a highly stable
tricapped trigonal prism unit which is seen to play a crucial role in the
finite temperature behavior of these clusters.  Thermodynamics of small
tin clusters (Sn$_{10}$ and Sn$_{20}$) is revisited in light of the
recent experiments on tin clusters of sizes 18-21 [G. A. Breaux~{\it et.
al.} Phys. Rev. B {\bf 71} 073410 (2005)].  We have calculated heat
capacities using multiple histogram technique for Si$_{10}$, Sn$_{10}$
and Si$_{15}$ clusters.  Our calculated specific heat curves have a main
peak around 2300~K and 2200~K for Si$_{10}$ and Sn$_{10}$ clusters
respectively. However, various other melting indicators such as root mean
square bond length fluctuations, mean square displacements show that
diffusive motion of atoms within the cluster begins around 650~K. The
finite temperature behavior of Si$_{10}$ and Sn$_{10}$ is dominated by
isomerization and it is rather difficult to discern the temperature range
for transition region.  On the other hand, Si$_{15}$ does show a liquid
like behavior over a short temperature range followed by the
fragmentation observed around 1800~K.  Finite temperature behavior of
Si$_{20}$ and Sn$_{20}$ show that these clusters do not melt but fragment
around 1200~K and 650~K respectively.  

\end{abstract}


\maketitle

\section{Introduction}
\label{sec.intro}

The finite temperature behavior of small clusters is of great interest
owing to the recent exciting calorimetric measurements carried out on
clusters of sodium, tin, gallium and aluminum~\cite{Hab-Na1, Jar-tin1,
Ga-PRL, Jar-jacs, Jar-Al}.  Calorimetric experiments on tin and gallium
clusters with 10-50 atoms reveal many interesting features. For example,
contrary to the standard paradigm, it has been shown that clusters can
have higher than bulk melting temperatures~\cite{Jar-tin1, Ga-PRL}.
Further, addition of a single atom is seen to change the specific heat
curve characteristics dramatically by changing a `magic melter' into a so
called `non-melter'~\cite{Jar-jacs}.  These experiments have motivated
many researchers to simulate the finite temperature behavior of small
clusters~\cite{PRB-Lu, Na-AMV, Our-PRBsn10, Our-PRBsn20, Our-PRL,
Eur-Phys.J, James-Tin, Our-Na}.

In the present work, we study the finite temperature behavior of small
silicon and tin clusters.  The silicon clusters are of potential
relevance to nanoelectronics industry and hence remain subject of many
experimental as well as theoretical
studies~\cite{PRL-spectroscopic-Muller, Jar-Chem-Soc, Photo, Si-Disso,
Jar-Nat, Si-JCP2, Si-laser, TTP, Si-PRL-Dynamics, Jackson, Si6, Si6-19,
Jackson-2004, Si-36, Si-JCP, Si-JCP1, Si-ACIE}. Most of the theoretical
studies are devoted to the investigation of lowest equilibrium geometries
of small silicon clusters, motivated by the experimental findings that Si
clusters undergo structural transition (prolate to spherical) in the
range 22-34~\cite{PRL-spectroscopic-Muller,Photo}. To date, global minima
of silicon clusters with sizes n$<$24 have been well established from
various recent unbiased searches such as genetic algorithm, single-parent
evolution algorithm, `big bang' optimization etc~\cite{Jar-Nat, Jackson,
Jackson-2004}.

The surface induced dissociation studies revealed that the fragmentation
behavior is common within the clusters of semiconducting group 14
elements~\cite{Tin-pathway, Ge-pathway}.  Studies on silicon clusters of
various sizes report existence of fragmentation pathway as compared to
evaporation process~\cite{Si-Disso, Si-pathway} and predict that clusters
with sizes 6-10 are the most abundant fragments and are referred to as
magic fragments~\cite{Si-PRL-Dynamics, Zhang-magic-jcp}.  Although, it is
speculated that silicon clusters up to sizes $70$ will undergo
fragmentation rather than evaporation~\cite{Si-Disso}, it is not yet
clear if the clusters undergo the traditional solid-like to liquid-like
transition before fragmenting or they fragment without melting as
reported recently in case of tin clusters~\cite{Jar-Tin-frag}.  Hence, we
report detailed thermodynamic simulations on Si$_{n}$ clusters with
various sizes; $n = 10,15,20$. These simulations have been carried out
using Density Functional Theory (DFT) within the Generalized Gradient
Approximation (GGA). The total simulation time for each cluster is at
least 2~ns.  Our detailed analysis of finite temperature behavior shows
that the finite temperature behavior of Si$_{10}$ is dominated with
isomerization and it is rather difficult to identify the region
corresponding to solid-like to liquid-like transition. However, the
cluster clearly fragments around 2800~K.  Si$_{15}$, on the other hand,
does exhibit a liquid like phase over a short temperature range before
fragmenting around 1800~K. Si$_{20}$ does not melt but fragments around
1200~K.

Both silicon and tin are group IV elements and it is interesting to note
some peculiarities of this group. Carbon, the first element is a
non-metal with a high energy gap between valence band and conduction band
(around 5.5~eV) whereas silicon and germanium are semiconductors at room
temperature. Tin and lead are metals at room temperature of which tin
undergoes a structural phase transition and transforms into a
semiconductor below 286~K.  Contrary to the bulk, tin clusters (n$<$25)
resemble greatly to that of small silicon and germanium
clusters~\cite{PRB-Lu, Jar-PRA-Tin-Prolate}.  Interestingly, Si, Ge and
Sn clusters with sizes $>$ 9 are built on stable TTP unit which, as we
shall see, plays a crucial role in finite temperature behavior of these
clusters. However, in contrast to silicon clusters, tin clusters undergo
a size rearrangement from prolate (stacked TTP units) to a more compact
spherical shape in a considerably broad range (from 35-65 atom cluster).
The ionic mobility experiments exploited the fact that small tin clusters
have prolate ground states.  Experiments on small tin clusters (10-30
atom clusters)  indicated that these clusters do not melt at least 50~K
above the bulk melting temperature.~\cite{Jar-tin1} This conclusion is
based on the argument that presence of a liquid phase is identified with
the change in the shape from prolate to spherical and hence enhanced
ionic mobilities in the liquid like region~\cite{Jar-tin1}.  Density
functional simulations supported these findings showing that, indeed,
melting temperatures of small tin clusters are at least 1000~K higher
than that of T$_{m[bulk]}$~\cite{PRB-Lu, Our-PRBsn10, Our-PRBsn20,
James-Tin}.  Our previous calculations, within LDA showed that Sn$_{10}$
has substantially higher melting temperature, 2300~K whereas Sn$_{20}$ is
in a liquid-like phase after 1200K~\cite{Our-PRBsn10,Our-PRBsn20}.
However, the recent experiments on Sn$_{18}^{+}$, Sn$^{+}_{19}$, 
Sn$_{20}^{+}$ and Sn$_{21}^{+}$ demonstrated that these clusters do not
melt, but sublimate~\cite{Jar-Tin-frag} around 650~K.  In light of these
findings, we simulate the thermodynamics of tin clusters, Sn$_{10}$ and
Sn$_{20}$, using GGA functionals.  In the present work, simulations are
extended over much longer time scale, at least 90~ps per temperature as
compared to our earlier simulations of 40~ps per
temperature~\cite{Our-PRBsn10, Our-PRBsn20}. Our studies confirm the
fragmentation observed experimentally and bring out similarities between
small tin and silicon clusters in terms of structure and dynamics.

In what follows, we present computational details in Sec.~\ref{sec:comp}.  
Results of both tin and silicon clusters are given in Sec.~\ref{sec:rd}
We discuss various issues concerning the structure and dynamics of all
the clusters in the same section. We conclude our paper in
Sec.~\ref{sec.concl}.

\section{Computational Details\label{sec:comp}}

All the thermodynamic simulations are performed using Born--Oppenheimer
molecular dynamics based on Kohn-Sham formulation of DFT~\cite{KS}. We
have used Vanderbilt's ultrasoft pseudopotentials~\cite{uspp-vanderbilt}
within GGA, as implemented in \textsc{vasp} package~\cite{vasp} for all
the clusters.  Thermodynamic behavior of tin clusters is also studied
using LDA as implemented in \textsc{vasp}.  For all the calculations, we
use only $3s^2$ and $3p^2$--electrons as valence in case of Si. In case
of Sn, we use $5s^2$ and $5p^2$--electrons as valence, taking
$d$--electrons as a part of the ionic core. An energy cutoff of 13.84~Ry
and 9.77~Ry is used for the plane--wave expansion of Si and Sn wave
functions respectively, with a convergence in the total energy of the
order of 10$^-4$~eV. Cubic supercells of lengths $15$, $18$ and
$22~{\AA}$ are used for X$_{10}$, X$_{15}$ and X$_{20}$, where $X=Si,Sn$,
respectively.

The ground state and other equilibrium structures of all the clusters are
found by optimizing several structures chosen periodically from a high
temperature simulation. We analyze the bonding characteristics of the
ground state and few excited state isomers of the clusters using Electron
Localization Function (ELF)~\cite{elf-silvi}. This function is normalized
between zero and unity; a value of 1 represents a perfect localization of
the valence charge while the value for the uniform electron gas is $0.5$.
The locations of maxima of this function are called \emph{attractors},
since other points in space can be connected to them by maximum gradient
paths. The set of all such points in space that are attracted by a
maximum is defined to be the \emph{basin} of that attractor. Basin
formations are usually observed as the value of the ELF is lowered from
its maximum, at which there are as many basins as the number of atoms in
the system. Typically, existence of an isosurface or a basin along the
bonding region between two atoms at a high ELF value, say $\ge 0.7$,
signifies a localized bond in that region.

For examining the finite temperature behavior, ionic phase space of all
the clusters is sampled by isokinetic MD where kinetic energy is held
constant using velocity scaling. In case of X$_{10}$ ($X=Si,Sn$)  
clusters we split the total temperature range from 100-3000~K into 22
different temperatures.  In case of Si$_{15}$ and Si$_{20}$ clusters we
split the temperature range from 100-1800~K into 14 different
temperatures.  Each system is heated to desired temperature from previous
temperature slowly within a time scale of 15~ps. Molecular dynamics is
then simulated at each temperature for 30~ps after which it is
subsequently heated to the next temperature. The MD for each temperature
is then continued for an additional time scale of 90 ps so as to have
sufficiently large statistics. In case of Sn$_{20}$, the molecular
dynamics simulations were carried out within the LDA as well as GGA
approximations.  In case of GGA, we have studied the finite temperature
behavior of the cluster around 200~K, 350~K, 500~K and 650~K. In case of
LDA calculations the temperature range is split so to have the finite
temperature behavior of the cluster around 100~K, 250~K, 500~K, 650~K,
 800~K and 1000~K.

Following the MD simulation, classical ionic density of states of system
is extracted using multiple histogram technique and the details are found
in literature~\cite{MH}. Various other thermodynamic indicators such as
Mean Square Displacements (MSD) of atoms and Root-Mean Square bond-length
fluctuations (RMS-BLF, $\delta_{\rm rms}$) are computed. More technical
details concerning the extraction of thermodynamics averages, indicators
and computation of specific heat curve can be found in our earlier
paper~\cite{amv-review}.

\section{Results and Discussion\label{sec:rd}}
\subsection{Thermodynamics of Silicon clusters}
\begin{figure}
  \epsfxsize=0.5\textwidth
  \centerline{\epsfbox{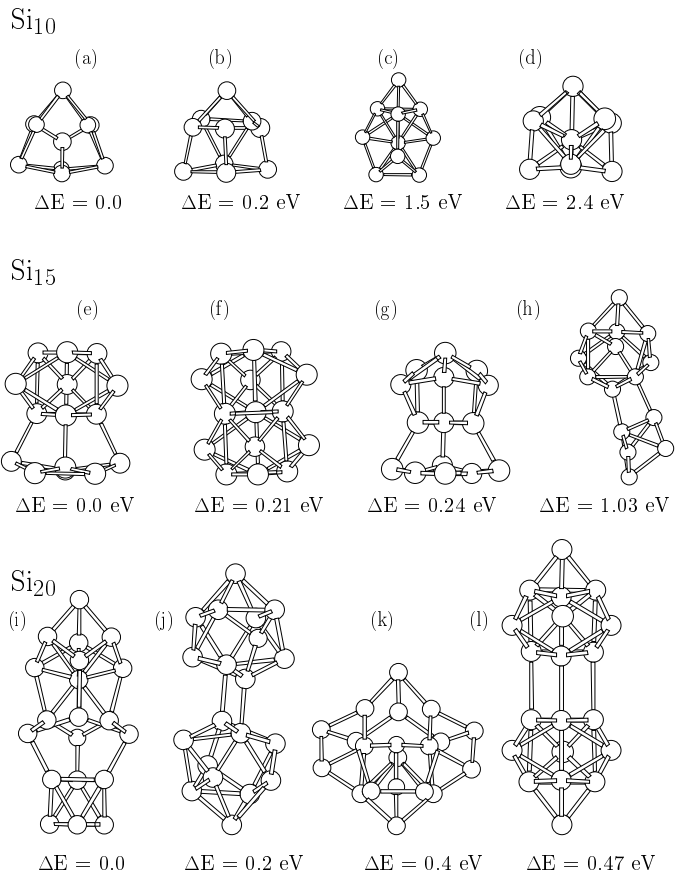}}
  \caption{\label{fig.geom}
  The ground state, and some representative excited
  states of Si$_{10}$, Si$_{15}$ and Si$_{20}$.
  The energy difference $\Delta E$ is given in eV with respect to the
  ground state.
  }
\end{figure}
First, we discuss the finite temperature behavior of Si$_{10}$. The
ground state and various excited state geometries of Si$_{10}$ are shown
in Fig.\ref{fig.geom} -- (a) to (d). The ground state of Si$_{10}$ is a
tetra capped trigonal prism (similar to that of
Sn$_{10}$~\cite{Our-PRBsn10}) and is consistent with the one reported
earlier~\cite{Si10geo}.  It may be noted with some interest that the
ground state and the first excited state structure (higher by 0.2~eV)
contain one TTP unit whereas all other excited configurations do not
contain a TTP unit. This explains the significant energy difference (of
$\approx$~1~eV) between the first two equilibrium structures and the
remaining excited state structures. We also note that in the ground state
geometry, all the atoms except one capping the triangular face of
trigonal prism have a minimum of four-fold coordination. The nature of
bonding is analyzed using ELF which shows the expected covalent bonding
among all the silicon atoms.

\begin{figure}
  \epsfxsize=0.5\textwidth
  \centerline{\epsfbox{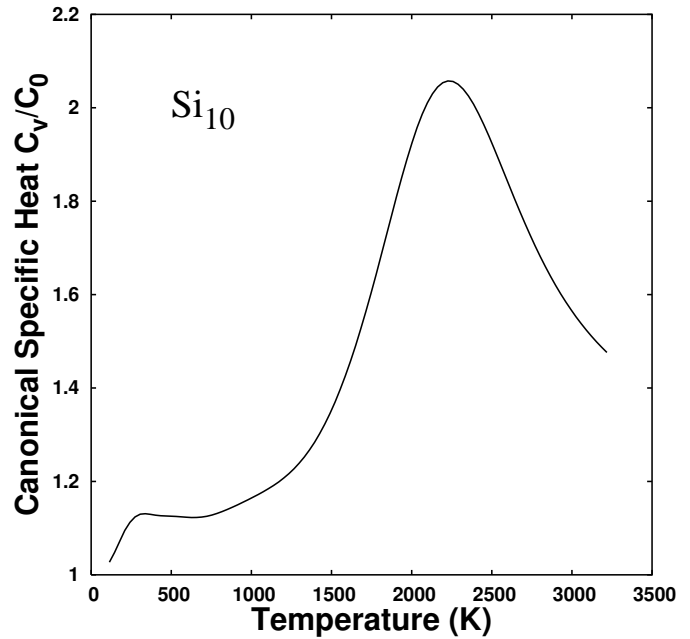}}
  \caption{\label{fig.cv.si10}
  Normalized canonical specific heat of Si$_{10}$.}
\end{figure}
\begin{figure}
  \epsfxsize=0.2\textwidth
  \centerline{\epsfbox{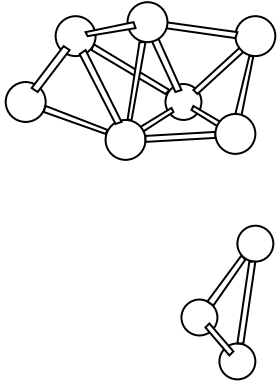}}
  \caption{\label{fig.disso.si10}
  Si$_{10}$ seen as fragmenting into Si$_{7}$ and Si$_{3}$ 
  at 2800K
  }
\end{figure}

Next, we discuss the calculated ionic specific heat curve for Si$_{10}$
shown in Fig.~\ref{fig.cv.si10}. The salient features of the curve are
presence of a broad shoulder from 350~K to 1400~K and a main peak at
2300~K. An analysis of ionic motion at various temperatures, reveals
several additional aspects. The shoulder at 350~K arises from a
structural transition of cluster from the ground state to the first
excited state and back to the ground state. This process is due to the
attempt of the tri-coordinated atom (the atom capping the triangular face
of the trigonal prism) to acquire the forth neighbor. The isomerization
seen around 350~K is quite similar to that of
Sn$_{10}$~\cite{Our-PRBsn10} and occurs in such a way that at least three
atoms (cap on triangular face, rectangular face and one from the trigonal
prism)  exchange their positions without changing the geometry
considerably.  This transition occurs with increased frequency until
1400~K and leads to the diffusion of atoms through out the cluster
without considerable change in the geometry. It is only at 1600~K and
beyond that structures corresponding to the other excited states (which
are higher in energy by almost 1 eV) are observed. Thus, the rise in the
specific heat curve observed after 1400~K is associated with the
destruction of the TTP unit and occurrence of other excited states.  At
much higher temperatures, around 2300~K and above, we observe several
high lying configurations which can be thought of as two weakly bonded
clusters of smaller sizes, with energies above 2.5~eV with respect to the
ground state. The cluster eventually fragments around 2800~K into Si$_7$
and Si$_3$ as shown in Fig.~\ref{fig.disso.si10}.  

\begin{figure}
  \epsfxsize=0.5\textwidth
  \centerline{\epsfbox{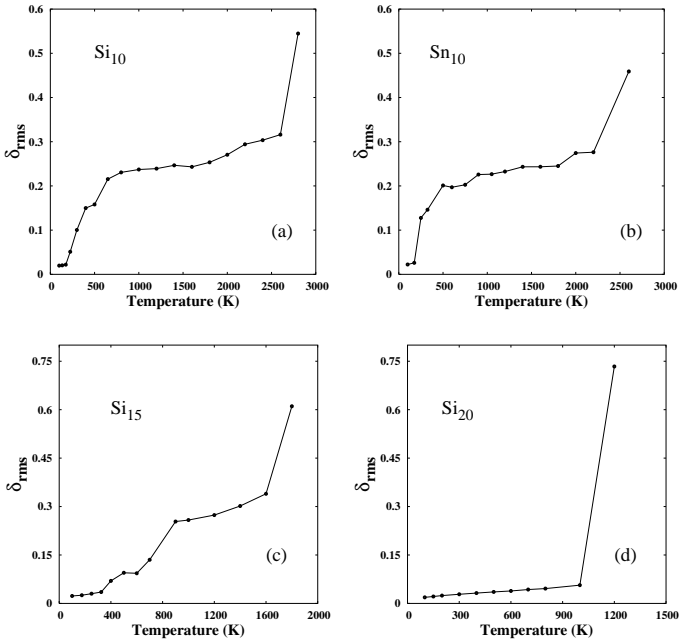}}
  \caption{\label{fig.delta}
  Root mean square bond length fluctuations ($\delta_{\rm rms}$)
  for Si$_{10}$, Sn$_{10}$, Si$_{15}$ and Si$_{20}$.}
\end{figure}

Thus, though the main peak in the specific heat curve of Si$_{10}$ is
seen around 2300~K, it cannot be attributed to the solid-like to
liquid-like transition. In fact, we believe that this peak is due to the
fragmentation that occurs around 2800~K. This is also evident from the
analysis of the root mean square bond-length fluctuations ($\delta_{\rm
rms}$) shown in Fig.~\ref{fig.delta}--(a). It is clearly seen that
$\delta_{\rm rms}$ is characterized by two sharp increments, first one
from 175-650~K and another at 2800~K.  According to the bulk Lindemann
criterion a value of $\delta_{\rm rms}$ greater than 0.1 is indicative of
transition from solid phase to
liquid phase. In bulk, the solid-liquid transition is sharp whereas in
clusters, the transition is gradual and is observed that, in a
liquid-like phase, the value of $\delta_{\rm rms}$ saturates around
$0.2-0.3$. At 650~K $\delta_{\rm rms}$ reaches a value of 0.2. This is
due to the fact that GS-First Excited state-GS transition results in
diffusion of atoms through out the cluster thereby saturating the value
of $\delta_{\rm rms}$. The value of $\delta_{\rm rms}$ is almost constant
for next 1000~K (i.e. up to 1600~K). Above 1600~K, coincident with the
destruction of TTP unit, the $\delta_{\rm rms}$ experiences a gradual and
slow rise until 2600~K. This gradual rise ends with a second sharp jump
at 2800~K due to fragmentation of the cluster.

\begin{figure}
  \epsfxsize=0.5\textwidth
  \centerline{\epsfbox{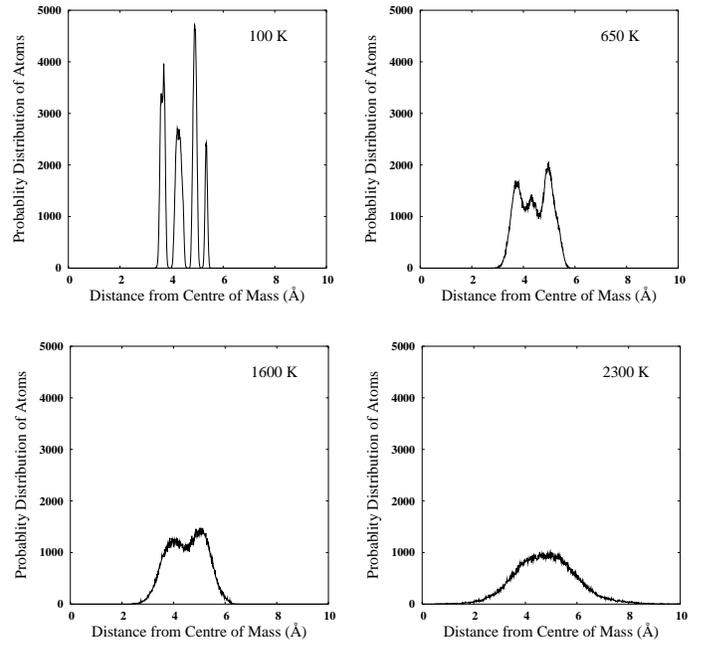}}
  \caption{\label{fig.atomic.distribution}
Atomic distribution from the centre of mass for Si$_{10}$ }
\end{figure}

Next we analyze the average distribution of atoms from Center of Mass
(COM) of the cluster as a function of temperature (shown in
Fig.~\ref{fig.atomic.distribution}). We have plotted the probability of
finding an atom as a function of distance from the COM. At low
temperatures, a perfect shell structure is retained, as observed for
100~K (shown in Fig.\ \ref{fig.atomic.distribution}--a).  It can be
easily identified that the first two peaks correspond to
the two faces of the trigonal prism, the third peak corresponds to the
caps on the rectangular surfaces and the last one is due to the cap on
the triangular face.  With increasing temperature, the peaks begin to
merge in the bottom, (as seen around and above 350~K) indicating an
exchange of atoms between various shells. Around 650~K, four peaks merge
into two and the distribution of atoms remains identical till 1400~K.  
Above 1600~K, the COM plot indicates absence of any sort of average order
in the arrangement of the atoms. The peak gets more broader at higher
temperatures and around 2300~K (coincident with the peak in the specific
heat), the atomic distribution from COM is seen as a broad curve.

In short, the main peak in the specific heat curve occurs around 2300~K
which is due to fragmentation. The Lindemann criteria, atomic
distribution of atoms around COM and other qualitative parameters
indicate diffusive motion of atoms within the cluster around 650~K.
Although, for a very broad range of temperatures, from 650~K to 1400~K,
the cluster undergoes characteristic isomerization discussed earlier, it
retains the shape of the cluster, but allows diffusion of the atoms
through out the cluster. Thus, finite temperature behavior of very small
clusters like Si$_{10}$, is dominated by isomerization.  The solid-like
to liquid-like transition is continuous and it is rather difficult to
discern the transition temperatures for the same.  This is also confirmed
by our on going studies on small sodium and lithium
clusters~\cite{Na-Mal-Soon}.

Turning to the finite temperature study of Si$_{15}$, we first discuss
the equilibrium structures of Si$_{15}$ shown in Fig.\ \ref{fig.geom}--(e
to h). The ground state geometry of Si$_{15}$ can be described as a
tricapped trigonal prism fused with a tricapped trigonal anti prism
(C$_{3v}$ symmetry Fig.\ \ref{fig.geom}--(e)). This is consistent with
the lowest energy configuration reported by CCSD(T)~\cite{Si-JCP} and DFT
simulations~\cite{Jackson}.  The first excited state consists of two TTP
units sharing a common triangular face as shown in Fig.\
\ref{fig.geom}--(f). This structure can be viewed as a serious distortion
of the lower six atom ring present below the TTP unit in the ground
state. The first few excited states are built on TTP unit of Si$_{9}$ as
shown in Fig.~\ref{fig.geom}--(f,g), whereas isomers observed at higher
temperatures can be thought of as a combination of two small silicon
clusters as shown in Fig.\ \ref{fig.geom}--(h) and suggest probable paths
for fragmentation.

\begin{figure}
  \epsfxsize=0.5\textwidth
  \centerline{\epsfbox{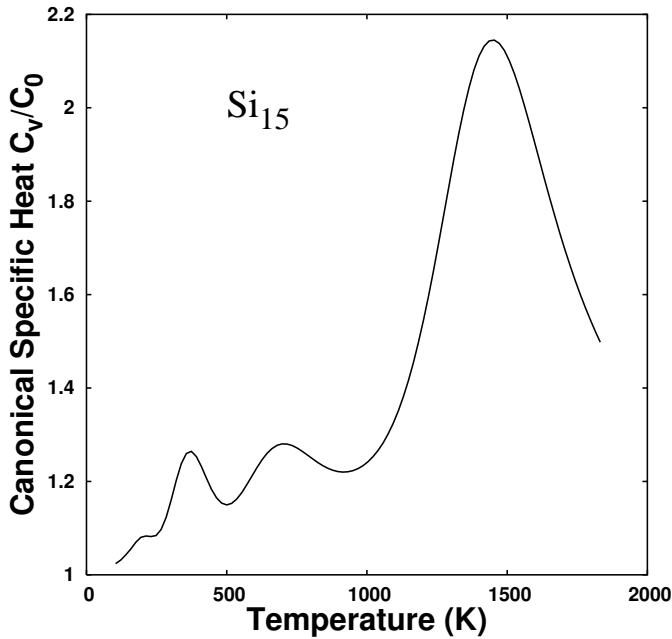}}
  \caption{\label{fig.cv.si15}
    Normalized canonical specific heat of Si$_{15}$.}
\end{figure}

The ionic specific heat curve computed for Si$_{15}$ (see Fig.\
\ref{fig.cv.si15})  has two small peaks around 400~K and 800~K and a main
peak around 1400~K which is associated with fragmentation of the cluster
around 1800~K.  Ionic motions of Si$_{15}$ cluster at various
temperatures provide a plausible explanation for different features
observed in the specific heat curve.
We would also like to note that Si$_{15}$ has many low lying isomers (we
have at least 30 different equilibrium structures out of which few
relevant structures are shown in the figure) unlike Si$_{10}$. The
isomerization begins around 400~K when we observe the first excited state
within our calculation (Fig.\ \ref{fig.geom}--(f)).  This involves
rearrangement of lower six atom ring.  With further rise in temperature,
the cluster remains in the first excited state for a reasonable amount of
time after which the atoms in upper TTP unit reorient into a six atom
ring leading to an inverted ground state. Thus, the first excited state
appears to be an intermediate for interchanging the positions of TTP unit
and the hexagonal ring in the ground state. This isomerization process
does not involve diffusion of atoms within the cluster and continues to
occur with increased frequency until 700~K. Further rise in temperature
leads to several other isomerization routes along with the one described
above which results in to diffusion of atoms through out the cluster.
This corresponds to the second shoulder seen in the specific heat curve
around 800~K (see Fig.\ \ref{fig.cv.si15}).  Around 1200~K isomers
corresponding to two small silicon clusters (Fig.\ \ref{fig.geom}--(h))
are observed in the ionic motion which provide a fragmentation path at
higher temperatures.

The $\delta_{\rm rms}$ (shown in Fig.\ \ref{fig.delta}--(c))  and the
distribution of atoms from COM of Si$_{15}$
\begin{figure}
 \epsfxsize=0.5\textwidth
 \centerline{\epsfbox{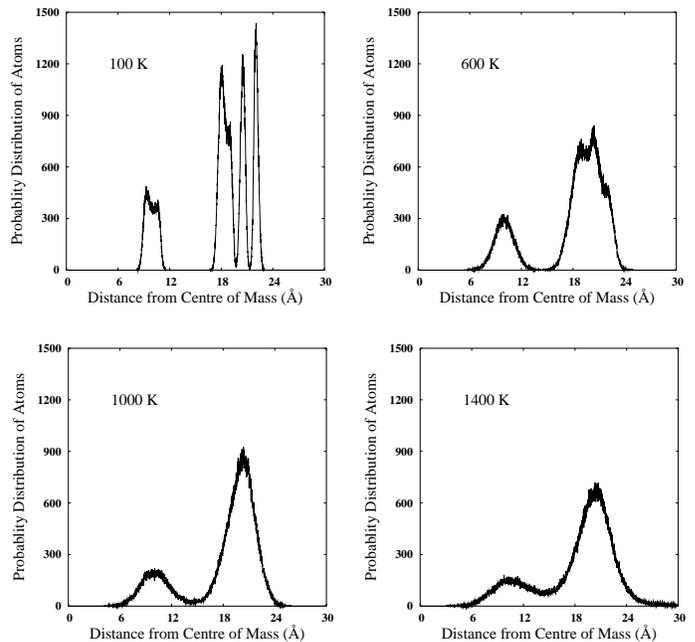}}
 \caption{\label{fig.com.si15}
 Atomic distribution from the center of mass for Si$_{15}$ }
\end{figure}
(shown in Fig.\ \ref{fig.com.si15}) support these observations. Linear
rise in $\delta_{\rm rms}$ at lower temperatures is indicative of pure
vibrational motion. Beginning of isomerization is shown up as a rise in
$\delta_{\rm rms}$ around 400~K which continues till 900~K. The value of
$\delta_{\rm rms}$ saturates after 900~K showing a liquid-like state over
a short temperature range.  Around this temperature, the diffusive motion
of atoms between the upper and lower units of Si$_{15}$ is clearly seen
in the distribution of atoms from the COM plot.  The value of
$\delta_{\rm rms}$ increases abruptly around 1800~K due to the
fragmentation of Si$_{15}$ into Si$_{9}$ and Si$_{6}$ clusters.

It is interesting to compare the finite temperature behavior of Si$_{15}$
and Si$_{10}$. Although, both clusters fragment, Si$_{15}$ (1800~K)  
dissociates at considerably lower temperature than Si$_{10}$
(2600~K)~\cite{note}. In case of Si$_{10}$, all isomers except one, are
at much higher in energy. As a result, the cluster undergoes a peculiar
isomerization (described earlier) over a wide range of temperatures
(350~K to 1400~K) leading to a broad shoulder in the specific heat curve
and a flat region in the $\delta_{\rm rms}$ plot. On the other hand, in
case of Si$_{15}$, the isomerization region is spread over comparatively
shorter range of temperatures (400~K to 900~K).

As we have already noted, for the clusters of group IV elements, TTP
forms a basic block on which larger clusters are built upon. As will be
discussed, the ground state of Si$_{20}$ is a prolate structure whereas
the TTP unit is evident in the excited states. Though the fragmentation
of Si$_{20}$ is expected, it will be interesting to see if the
fragmentation is preceded by a liquid like state. In what follows, we
present the thermodynamics of Si$_{20}$ cluster. In Fig.\
\ref{fig.geom}--(i) to (l), we show the ground state as well as some of
the excited state geometries of Si$_{20}$.  The lowest-energy isomer of
Si$_{20}$ is composed of three units: a so called magic number cluster
Si$_{6}$, a hexagonal chair unit in the middle, and a low energy isomer
of Si$_{8}$ unit.  The first excited state consist of two distinct TTP
units whereas the second one is non-prolate, spherical structure.  The
forth equilibrium structure that we have shown in the figure, (see Fig.\
\ref{fig.geom}--(l)) has two Si$_{10}$ units connected via very weak
bonds (bond lengths $\approx 2.7 \AA$ as compared to the normal Si-Si
covalent bondlength of 2.34~{\AA}). As we shall see, the cluster
fragments via this isomer. All the equilibrium structures shown here are
consistent with earlier reports~\cite{Jackson-2004,Si-PRL-Dynamics}.

The nature of bonding is examined using ELF. As expected, ELF for the
ground state shows a high covalent bonding with all the basins completely
connected by an isovalue of 0.75. It is also observed from ELF that the
two TTP units in Fig.\ \ref{fig.geom}--(l) are weakly bonded to each
other. Turning to the finite temperature behavior of Si$_{20}$, we note
that till 800~K, atoms in the cluster vibrate about their mean positions
and no isomerization is observed. Around 1000~K, the cluster transforms
from the ground state to the excited state having two distinct Si$_{10}$
units (Fig.\ \ref{fig.geom}-(l)) and eventually fragments into two
Si$_{10}$ units around 1200~K. This is reflected in the $\delta_{\rm
rms}$ shown in Fig.\ \ref{fig.delta}--(d) where the fragmentation is
evident from the sharp rise around 1200~K. In other words, finite
temperature behavior of Si$_{20}$ does not show solid-liquid transition
prior to fragmentation. This may be attributed to the low dissociation
energy barrier in case of Si$_{20}$ cluster (of the order of 1.2~eV) in
comparison to that of Si$_{10}$ and Si$_{15}$ (which have a dissociation
energy barrier of 4.2~eV and 2.2~eV respectively)~\cite{Si-Disso}.  To
understand the influence of starting geometry on the fragmentation
behavior of the cluster, we have repeated the thermodynamic simulations
on Si$_{20}$ cluster with Fig.\ \ref{fig.geom}--(j) and Fig.\
\ref{fig.geom}--(k) as the starting configurations. We observe the
cluster to fragment around 1200~K in both these calculations as well
indicating that the fragmentation behavior is not sensitive to the
starting configuration of the cluster.

We end our discussion on silicon clusters with an interesting
observation. The observed fragmentation can be bypassed with higher rate
of heating.  In all these simulations, the cluster is heated to a
particular temperature from the previous temperature at a rate of 1 K in
30~fs and then maintained at that temperature for next 30~ps. This is
used as a starting point for heating the cluster to the next temperature.
We observed the fragmentation after maintaining the cluster at 1200~K for
at least 60~ps. On the other hand when the cluster is heated from 1200~K
to 1600~K, without sufficiently thermalizing it at 1200~K, it shows
liquid-like behavior at 1600~K. This liquid like state is also observed
for 1800~K~\cite{note1}.
\begin{figure}
  \epsfxsize=0.4\textwidth
  \centerline{\epsfbox{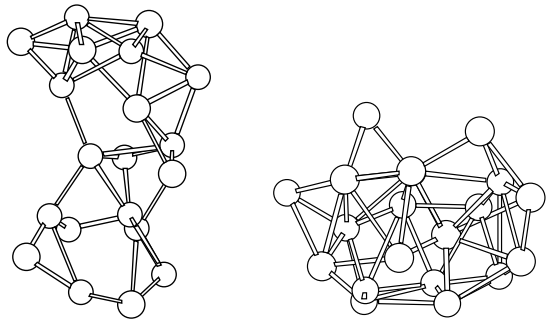}}
  \caption{\label{fig.config}
    Prolate and spherical configurations of Si$_{20}$ around 1800~K}
\end{figure}
It is seen that while heating the cluster without allowing enough time
for thermalization, it attains enough kinetic energy to push the two
Si$_{10}$ units closer to each other (Si$_{TTP1}$-Si$_{TTP2}$ bond
distance of 2.2~\AA). Consequently, the two TTP units begin to interact
leading to high lying configurations (containing no TTP units) as shown
in Fig.~\ref{fig.config}. The ionic motion of this super heated cluster
shows that the cluster does not fragment at the end of 100~ps. On the
contrary, a complete diffusive motion of atoms within the cluster is
observed.
\begin{figure}
  \epsfxsize=0.35\textwidth
  \centerline{\epsfbox{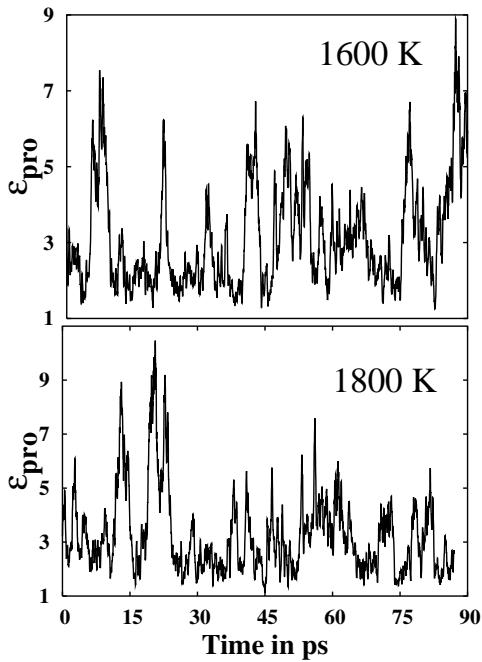}}
  \caption{\label{fig.shape}
Deformation coefficient, $\epsilon_{pro}$, as a function of simulation
time for Si$_{20}$ cluster}
\end{figure}
The temperature of the cluster was further raised to 1800~K.  The cluster
was then maintained at that temperature for an additional time scale of
100~ps. An analysis of ionic motion at this temperature shows the cluster
to be in a liquid-like state there by revealing a possibility to bypass
the fragmentation process. An examination of ionic motion at 1600~K and
1800~K reveals many interesting features. Around 1600~K, the cluster
oscillates between prolate configurations and non-prolate (near
spherical) configurations several times. As the temperature rises, the
cluster spends more time within the near spherical configurations. This
shape change can be experimentally detected using ion mobility
experiments (diffusion coefficient)~\cite{Jar-tin1}. Hence, is it
interesting to analyze these shape changes in a more detailed fashion. In
Fig.\ \ref{fig.shape}, we have plotted the deformation coefficient
($\epsilon_{\rm pro}$)  of the cluster as a function of time, at 1600~K
and 1800~K respectively. For a given ionic configuration, $\epsilon_{\rm
pro}$ is defined as 
\begin{equation} 
\epsilon_{\rm pro} = \frac{2Q_1}{Q_2+Q_3} 
\end{equation} 
where, Q$_{1}$,Q$_{2}$ and Q$_{3}$
are the eigenvalues, in descending order of the quadrapole tensor
\begin{equation} 
Q_{ij}=\sum_{I}R_{Ii}R_{Ij} 
\end{equation} 
I runs over
the number of ions and R$_{Ii}$ is the ith coordinate (i and j run from 1
to 3) of the ion `I' relative to the center of mass of the cluster. An
ionic configuration with spherical shape has a value of $\epsilon_{pro}$
= 1. Deviation of the value of from 1 is a measure of prolaticity of the
cluster. As is seen from the figure, the cluster oscillates between
prolate and nearly spherical configurations quite frequently around
1600~K. At 1800~K, the oscillations reduce and the structure is seen to
visit the spherical configurations more often. The value of $\delta_{\rm
rms}$ at 1600~K and 1800~K is 0.35 and 0.37 respectively and satisfies
the Lindemann criterion.

\subsection{Thermodynamics of Tin clusters}

Various interesting observations seen in case of the silicon clusters and
recent experiments~\cite{Jar-Tin-frag} on tin clusters (Sn$_{19}$,
Sn$_{19}$, Sn$_{20}$ and Sn$_{21}$) have motivated us to revisit the
thermodynamics of tin clusters (Sn$_{10}$ and Sn$_{20}$).  In the present
work, the finite temperature study is carried out within the GGA as well
as LDA approximations using the vasp package~\cite{vasp}.  The clusters
are heated to each temperature at a much slower rate (1~K in 0.036 ps)  
than in our earlier work. The cluster was maintained at each temperature
for a period of aleast 90~ps (as contrast to 40~ps per temperature in our
earlier work). The heat capacity curves are computed from the above
calculations (GGA and LDA) and compared with our earlier results.

The finite temperature behavior of Sn$_{10}$ is seen to be considerably
similar to that of Si$_{10}$. This is in fact expected result as the
ground state and even the excited state configurations of Si$_{10}$ and
Sn$_{10}$ are identical.  It is gratifying to note that the specific heat
curves computed from GGA as well as LDA calculations are nearly similar
to that of our earlier reported one ~\cite{Our-PRBsn10} (with a slightly
shifted main peak around 2200~K as compared to our earlier reported one
at 2300~K). Various inferences concerning bonding and ionic motion are in
perfect agreement with our earlier report. The significant difference to
be noted is the fragmentation of Sn$_{10}$ observed at 2600~K. This
process is observed only after a time scale of ~60~ps.  We believe this
feature to be missing in our earlier reports due to the smaller
simulation time scales (~50~ps per temperature).  However, it may be
recalled, that ionic motion of cluster at 2600~K in our earlier study
revealed few configurations corresponding to weakly interacting Sn$_{x}$
and Sn$_{10-x}$ units.  Thus, we conclude that the peak in the specific
heat curve is due to the fragmentation of Sn$_{10}$ cluster seen at
2600~K and not due to the traditional solid-liquid transition as
reported.

We have also computed the ionic specific heat of Sn$_{20}$ cluster within
GGA and LDA approximations. Analysis of the ionic motion of Sn$_{20}$
cluster (GGA) at various temperatures shows that the cluster vibrates
around its ground state until 600~K. Around 650~K, the cluster fragments
into two TTP (Sn$_{10}$) units. This is in perfect agreement with the
recently reported experimental observations ~\cite{Jar-Tin-frag}. Finite
temperature study of Sn$_{20}$ cluster within LDA approximation shows the
cluster to fragment into two Sn$_{10}$ units around 1000~K. The failure
to detect the fragmentation behavior in our earlier reports is attributed
to the following reasons: (a) faster heating rate (1~K in 0.01 ps)  with
smaller simulation time scales in our earlier simulations (40~ps per
temperature).  (b) Use of LDA functional in our earlier reports. As
already mentioned ground state Sn$_{20}$~\cite{Our-PRBsn20} cluster has
inhomogeneous bond strength distribution. While the bonds within
intra-TTP atoms are strongly covalent in nature, inter-TTP atoms are
connected through very weak covalent bonds. The LDA approximation was
unable to distinguish these differences leading to an over-estimation of
binding energies between the two TTP units. Hence, the Sn$_{20}$ cluster
did not fragment in the small simulation times scales reported earlier.
Sn$_{10}$ cluster in contrast, has all its atoms connected with similar
bond strengths, resulting in a more accurate finite temperature behavior
within the LDA approximation.

\section{conclusions}
\label{sec.concl}

We have presented thermodynamics of small silicon and tin clusters
(Si$_{10}$, Si$_{15}$, Si$_{20}$, Sn$_{10}$ and Sn$_{20}$) within LDA and
GGA approximations of DFT. Finite temperature behavior of Si$_{10}$ and
Sn$_{10}$ is dominated by isomerization process and fragmentation is
observed around 2800~K and 2600~K respectively.  The peak in the specific
heat curves around 2300~K and 2200~K for Si$_{10}$ and Sn$_{10}$
correspond to the observed fragmentation rather than solid-like to
liquid-like transition. The similarities observed in the finite
temperature behavior of these two clusters suggest the influence of
ground state structures on the finite temperature behavior of a cluster.
The specific heat curve of Si$_{15}$ has a main peak around 1400~K which
is due to the fragmentation observed around 1600~K. However, the cluster
exhibits a liquid like behavior over a short temperature range (900~K to
1400~K) before fragmenting. Si$_{20}$ and Sn$_{20}$ are found to
dissociate without melting around 1200~K and 650~K respectively. Our
simulations on Sn$_{20}$ agree with the recently reported experimental
results.

\section{acknowledgments}

We are grateful to Prof. Jackson for providing us with the ground state
configuration of Si$_{20}$. C--DAC (Pune) is acknowledged for providing
us with the supercomputing facilities. The authors thank IFCPAR for
partial financial support.


\begin{thebibliography}{}

\bibitem{Hab-Na1}     M. Schmidt, R. Kusche, B. von
                      Issendorff, and H. Haberland,
                      Nature (London) {\bf 393}, 238 (1998).
                      M. Schmidt and H. Haberland,
                      C. R. Physique {\bf 3} 327, (2002),
                     H. Haberland, T. Hippler, J. Dongres, O. Kostko,
                     M. Schmidt, and B. von Issendroff.
                      Phys. Rev. Lett. {\bf 94} 035701 (2005).


\bibitem{Jar-tin1}    A. Shvartsburg and M. F. Jarrold, Phys. 
                      Rev. Lett. {\bf 85}, 2530 (2000)

\bibitem{Ga-PRL}      G. A. Breaux, R. C. Benirschke, T. Sugai,
                        B. S. Kinnear, and M. F. Jarrold,
                        Phys. Rev. Lett. {\bf 91}, 215508 (2003).

\bibitem{Jar-jacs}      G. A. Breaux, D. A. Hillman, C. M. Neal,
                        R. C. Benirschke, and M. F. Jarrold,
                        J. Am. Chem. Soc. {\bf 126}, 8628 (2004).

\bibitem{Jar-Al}      G. A. Breaux, C. M. Neal, B. Cao, and M.F. Jarrold
                      Phys. Rev. Lett. {\bf 94}, 173401 (2005).

\bibitem{PRB-Lu}        Z.Y. Lu, C.Z. Wang, and K.M. Ho, 
                        Phys. Rev. B {\bf 61}, 2329 (2000)

\bibitem{Na-AMV}     A. M. Vichare,  D.G. Kanhere, and  S. A. Blundell 
                     Phys. Rev. B {\bf 64} 045408 (2001) 

\bibitem{Our-PRBsn10}  K. Joshi, D. G. Kanhere, and S.A. Blundell,
                       Phys. Rev. B {\bf 66} 155329 (2002)

\bibitem{Our-PRBsn20}  K. Joshi, D. G. Kanhere, and S.A. Blundell,
                       Phys. Rev. B {\bf 67} 235413 (2003)

\bibitem{Our-PRL}     S. Chacko, Kavita Joshi, D.G. Kanhere, and
                      S. A. Blundell, Phys. Rev. Lett.
                       {\bf 92}, 135506 (2004)

\bibitem{Eur-Phys.J}  M.Manninen, A. Rytkonen, and M. Manninen Eur.
                      Phys. J. {\bf 29}, 39 (2004)

\bibitem{James-Tin}    F.C. Chuang, C.Z. Wang, S.Ogut, J.R.
                       Chelikowsky, and K.M. Ho,
                       Phys. Rev. B {\bf 69}, 165408 (2004)

\bibitem{Our-Na}     S. Chacko, D.G. Kanhere, and S. A. Blundell,
                     Phys. Rev. B {\bf 71}, 155407 (2005) 

\bibitem{Photo}  K. Fuke, K. Tsukamoto, F. Misaizu, and M. Sanekata,
               J. Chem. Phys. {\bf 99}, 7807 1993.

\bibitem{Si-Disso} A.A. Shvartsburg, M.F. Jarrold, B. Liu, Z.Y. Lu, 
                   C.Z. Wang, and K-M. Ho
                    Phys. Rev. Lett. {\bf 81}, 4616 (1998)

\bibitem{Jar-Nat}     K.-M. Ho, A.A. Shvartsburg, B. Pan, Z.Y. Lu, C.Z
                      Wang, J.G. Wacker, J.L. Fye, and M.F. Jarrold,  
                      Nature, {\bf 392}, 582 (1998)

\bibitem{Si-JCP2}   B. Liu, Z.Y. Lu, B. Pan, C.Z. Wang, K.M. Ho,
                    A.A. Shvartsburg, and M.F. Jarrold
                    J. Chem. Phys. {\bf 109}, 9401 (1998)

\bibitem{Si-laser}   K. Zickfeld, M.E. Garcia, and K.H. Bennemann
                    Phys. Rev. B {\bf 59}, 13422 (1999)

\bibitem{TTP}         A.A. Shvartsburg and M.F. Jarrold, Chem. Phys.
                      Lett. {\bf 317}, 615 (2000)


\bibitem{PRL-spectroscopic-Muller}   J. Muller, B. Liu, A.A. Shavartsburg, S. Ogut, 
                                     J. Chelikowsky, K.W.M. Sui, K-M. Ho, and G. Gantefor 
                                     Phys. Rev. Lett.  {\bf 85}, 1666 (2000)

\bibitem{Si-PRL-Dynamics} L. Mitas, J.C. Grossman, I. Stich, and J. Tobik                        
                          Phys. Rev. Lett. {\bf 84}, 1479 (2000)

\bibitem{Jackson}   I. Rata, A. A. Shvartsburg, M. Horoi, T.Frauenheim,
                    K. W. Michael Siu, and K. A. Jackson
                    Phys. Rev. Lett. {\bf 85}, 546 (2000).

\bibitem{Si6}       A.D. Zdetsis
                    Phys. Rev. A {\bf 64}, 023202 (2001)

\bibitem{Jar-Chem-Soc}  A.A. Shvartsburg, R.R. Hudgins, P. Dugourd, and M.F. Jarrold
                        Chem. Soc. Rev. {\bf 30}, 36 (2001)

\bibitem{Si6-19}     B.K. Panda, S. Mukherjee, and S.N. Behera
                    Phys. Rev. B {\bf 63}, 045404 (2001)

\bibitem{Jackson-2004}    K.A. Jackson, M. Horoi, I. Chaudhuri,
                T. Frauenheim, and A. A. Shvartsburg  
                Phys. Rev. Lett. {\bf 93}, 013401, 2003.

\bibitem{Si-36}        Q. Sun, Q. Wang, P. Jena, S. Waterman, and Y. Kawazoe,
                       Phys. Rev. A {\bf 67}, 063201 (2003)

\bibitem{Si-JCP}    X. L. Zhu and X. C. Zeng
                    J. Chem. Phys. {\bf 118}, 3558 (2003)

\bibitem{Si-JCP1}    X. L. Zhu, X. C. Zeng, and B. Pan
                    J. Chem. Phys. {\bf 120}, 8985 (2004)

\bibitem{Si-ACIE}    S. Yoo and X.C. Zeng
                     Angew. Chem. Int. Ed. {\bf 44}, 1491 (2005)

\bibitem{Si-pathway}   Y.Tai and J. Murakami
                        Chem. Phys. Lett. {\bf 339}, 9 (2001)


\bibitem{Tin-pathway}   Y.Tai, J. Murakami, C. Majumdar, V. Kumar, H. Mizuseki, 
                        and Y. Kawazoa, 
                        J. Chem. Phys. {\bf 117}, 4317 (2002)

\bibitem{Ge-pathway}   Y.Tai, J. Murakami, C. Majumdar, V. Kumar, H. Mizuseki, 
                        and Y. Kawazoa, 
                        Eur. Phys. J. D {\bf 24}, 295  (2003)

\bibitem{Zhang-magic-jcp}  A.L. Zhang, Y. Liu, R.F. Curl, F.K. Tittel,
                           and R.E. Smalley, 
                           J. Chem. Phys. {\bf 88}, 1670 (1988) 

\bibitem{Jar-Tin-frag} G.A. Breaux, C. M. Neal, B. Cao, and M.F. Jarrold   
                        Phys. Rev. B {\bf 71}, 073410 (2005)

\bibitem{Jar-PRA-Tin-Prolate}   A.A. Shvartsburg and M.F. Jarrold, 
                               Phys. Rev. A {\bf 60}, 1235 (1999)

\bibitem{KS}      M.C. Payne, M.P. Teter, D.C. Allen, T.A. Arias, and
                  J. D. Joannopoulos,     
                  Rev. Mod. Phys. {\bf 64}, 1045 (1992).

\bibitem{uspp-vanderbilt}   D. Vanderbilt,
                            Phys. Rev. B {\bf 41}, 7892 (1990).

\bibitem{vasp}       Vienna {\em ab initio } simulation package,
                     Technische Universit\"at Wien (1999);
                     G. Kresse and J. Furthm\"uller,
                     Phys. Rev. B {\bf 54}, 11169 (1996).

\bibitem{elf-silvi}   B. Silvi, and A. Savin,
                      Nature (London), {\bf 371}, 683 (1994).

\bibitem{MH}       A. M. Ferrenberg and R. H. Swendsen,
                   Phys. Rev. Lett. {\bf 61}, 2635 (1988);
                   P. Labastie and R. L. Whetten,
                   Phys. Rev. Lett. {\bf 65}, 1567 (1990).

\bibitem{amv-review} D. G. Kanhere, A. Vichare, and S. A. Blundell,
                    {\em Reviews in Modern Quantum Chemistry},
                     Edited by K. D. Sen, World Scientific,
                     Singapore (2001).

\bibitem{Si10geo}  K. Raghavachari and C. M. Rohlfing, 
                   J. Chem. Phys. {\bf 89}, 2219 (1988).
                   P. Ballone, W. Andreoni, R. Car, and M. Parrinello, 
                   Phys.  Rev. Lett. {\bf 60}, 271 (1988).
                   C. M. Rohlfing and K. Raghavachari, 
                   Chem. Phys. Lett.  {\bf 167}, 559 (1990).

\bibitem{Na-Mal-Soon}     Mal-Soon Lee, S. Chacko, and D. G. Kanhere
                          (submitted to J. Chem. Phys.)

\bibitem{note} Fragmentation is first observed at 
               these temperarures and it is shown up
               as a sudden rise in the value of 
               $\delta_{\rm rms}$. See Fig.\ \ref{fig.delta} 

\bibitem{note1} The observed liquid-like behavior 
                sustains for atleast 100~ps.

\end{thebibliography}
\end{document}